\documentclass[referee]{raa}
\usepackage{graphicx,times}             
\usepackage{natbib}
\usepackage{amssymb,amsmath}
\bibpunct{(}{)}{;}{a}{}{,}

\usepackage[pagebackref=true]{hyperref}

\newcommand{\target}{{\rm MAXI~J1631-479\ }}

\newcommand{\NuSTAR}{{\it NuSTAR\ }}
\newcommand{\risco}{{$r_{\rm{ISCO}}$\ }}
\newcommand{\ergscm}{{${\rm erg~s^{-1} cm^{-2}}$}}
\newcommand{\ergcms}{{${\rm erg~cm~s^{-1} }$}}

\begin{document}

\title{The width-flux relation of the broad iron K$\alpha$ line during the state transitions of the black hole X-ray binaries}

\volnopage{Vol.0 (200x) No.0, 000--000}      
\setcounter{page}{1}          

\author{Hang-Ying Shui\inst{1,3}
\and Fu-Guo Xie \inst{2,3}
\and Zhen Yan \inst{2,3}
\and Ren-Yi Ma  \inst{1,3}}

\institute{Department of Astronomy and Institute of Theoretical Physics and Astrophysics, Xiamen University, Xiamen, Fujian 361005, China; {\it {ryma@xmu.edu.cn}} \\
\and Key Laboratory for Research in Galaxies and Cosmology, Shanghai Astronomical Observatory, Chinese Academy of Sciences, 80 Nandan Road, Shanghai 200030, China; {\it fgxie@shao.ac.cn}\\
\and SHAO-XMU Joint Center for Astrophysics, Xiamen, Fujian 361005, China}

\date{Received~~2009 month day; accepted~~2009~~month day}

\abstract{
The observation of varying broad iron lines during the state transition of the black hole X-ray binaries (BHXBs) have been accumulating.In this work, the relation between the normalized intensity and the width of iron lines is investigated, in order to understand better the variation of iron lines and possibly its connection to state transition. Considering the uncertainties due to ionization and illuminating X-rays, only the effects of geometry and gravity are taken into account. Three scenarios were studied, i.e., the continuous disk model, innermost annulus model, and the cloud model. As shown by our calculations, at given iron width, the line flux of the cloud model is smaller than that of the continuous disk model; while for the innermost annulus model, the width is almost unrelated with the flux. The range of the line strength depends on both the BH spin and the inclination of the disk. We then apply to the observation  of \target  by \NuSTAR during its decay from the soft state to the intermediate state. We estimated the relative line strength and width according to the spectral fitting results in \cite{Xu_2020}, and then compared with our theoretical width-flux relation. It was found that the cloud model was more favored.
We further modeled the iron line profiles, and found that the cloud model can explain both the line profile and its variation with reasonable parameters. 
\keywords{accretion, accretion disks -- black hole physics -- X-rays: binaries}
}

\authorrunning{Shui et al.}
\titlerunning{The Width-flux Relation of the Fe K$\alpha$ Line}
\maketitle

\section{Introduction}

As well known, the outbursts of the BHXBs usually go through different spectral states in sequence, i.e., from the quiescent state to the hard state (HS), the intermediate state (IS), the soft state (SS), and then back inverse \citep[e.g.][]{2006ARA&A..44...49R}. 

The theoretical models for the accretion flows in the HS and SS have been widely accepted.
For the SS, the X-ray spectrum is dominated by the soft thermal component, the hard power-law component is weak with the photon index larger than 2, and the accretion flow is believed to be in the mode of cold standard accretion disk (SAD) that extends to the innermost stable circular orbit (ISCO) \citep[e.g.][]{1973A&A....24..337S}, being sandwiched by rare hot corona \citep[e.g.][]{Liang1977}.
For the HS, the power-law component dominates the spectrum, and the accretion flow transforms to the hot accretion flow inside the truncation radius \citep[e.g.][]{Narayan1994,Yuan2014ARAA}.

However, the accretion mode of the IS, which describes how the accretion flow transits between the SAD and the hot accretion flow, is still unclear. Geometrically, there are three possible scenarios.
A natural scenario is the gradual inward extending or outward receding of the outer cold disk down to/from the ISCO when the accretion rate increases/decreases, or the continuous disk model, as first proposed by \cite{Esin1997}.

Additionally, considering the evaporation and condensing between the rare hot gas/corona and the dense cold disk, theoretical studies have shown that an annulus of dense cold gas may form or vanish in the inner region of the hot accretion flow during the state transition \citep{Liu2007,Liu2011,Qiao2011}. This kind of two-phase accretion flow is referred as the innermost annulus model in our paper, since the annulus is not very far from the ISCO.

In fact, the geometry of the two-phase accretion flow could be even more complex.
Instead of one annulus, the cold-phase gas may also be in form of clouds or clumps.
Such inhomogeous accretion flow could relates to some instabilities \citep{1996MNRAS.283.1322K,Blaes_2001,Blaes_2003,1998MNRAS.297..929G}.  {For example, if the advection heating could not balance the radiative cooling, the thermal instability would be triggered, and the luminous hot accretion flow would collapse non-linearly, leading to the formation of clumps} \citep{Yuan2001,Yuan2003,2012MNRAS.427.1580X}.
The existence of such cloudy accretion flow have been shown in some numerical simulations \citep{Wu2016,Bu2018,Sadowski2017,2022arXiv220103526L}.

However, from observational point of view, it is still difficult to identify the process of state transition.
On the one hand, the observations with high spectral resolution during the state transition are still lacking.
On the other hand, different observations may have different results.
In some observations, a second thermal component is needed to fit the spectrum, which seem to favor the cloud or the annulus model \citep[e.g.,][]{Walton_2017,Miller_2018,Tomsick_2018}.
While for some other observations, the truncation radius has already reached the ISCO before the hard-to-soft transition starts to occur \citep[e.g.][]{Parker_2015,2019MNRAS.490.1350B,2022MNRAS.514.1422D},
which seems to disfavor all the three scenarios mentioned above.

Recently, the state transition of the black hole X-ray binary \target has been observed by \NuSTAR, where the evolution of the broad iron line was obtained \citep{Xu_2020}.
 The authors fitted the spectra with the continuous disk model, and found that the truncation radius evolves from $< 1.9 ~r_g$ to $12\pm 1~r_g$. However, the authors found that the reflection rate in SS is usually high relative to that in IS, and the iron line flux is not correlated with the strength of the coronal emission in short intervals. They suggested that the thermal emission may also contribute to the line emission, however detailed calculation was not given.

The observed evolution of line width and strength enlightened us to study the possible relation between the flux and width of the broad iron. For the present spectral resolution, it is still difficult to identify the scenario of state transition by the fitting of iron line profile, as the fluctuation of the profile is still too large. The variation of the line flux and flux may provide us the clues of state transition for present observations.

The reflection features are related with a series of factors, such as the illuminating X-ray spectra and intensity, the ionization degree of the cold gas, the geometry and movement of the cold gas, and the gravitational field. Considering that the former two factors are still difficult to be constrained from theory and they vary for different outbursts and different stages of one burst, as the primary work, we just take into account the effects due to geometry and gravity. Our model is very simple, but the result is quantitative and robust.

In Sec.2, our model is introduced, where the geometry of the accretion flow are described by a set of three parameters. For different model of state transition, the parameters vary in different ways. In Sec.3, we investigate how the line strength changes with the line width for different models. All the line strength are normalized by the illuminating hard X-rays and the line flux in the soft state. The favored scenario for the state transition of \target are discussed in Sec.4, and the summary is given in Sec.5.

\begin{figure}
    \includegraphics[angle=-90,width=\columnwidth]{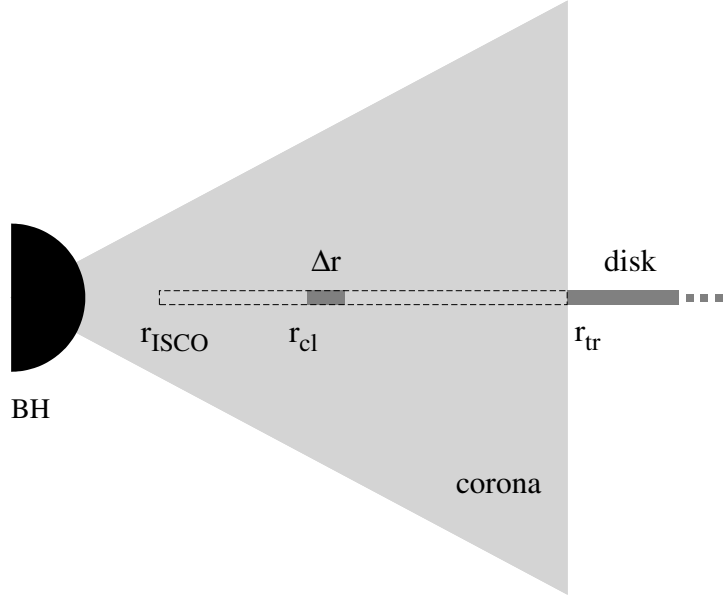}
    \caption{The geometry of the inhomogeneous accretion flow in our model. The dark grey rectangles denote the cold gas, i.e. disk and cloud. The light grey triangle denotes the hot accretion flow or corona.  {The dashed line shows the geometry of SAD in the inner region, which extends to ISCO, $r_{\rm{ISCO}}$}. The parameters $r_{\rm{cl}}$, $\Delta r$ and $r_{\rm{tr}}$ represent the inner edge and radial size of the clumps, and the radius of truncation, respectively.}
    \label{fgeom}
\end{figure}

\begin{figure}
\begin{center}
 	\includegraphics[angle=-90,width=0.8\columnwidth]{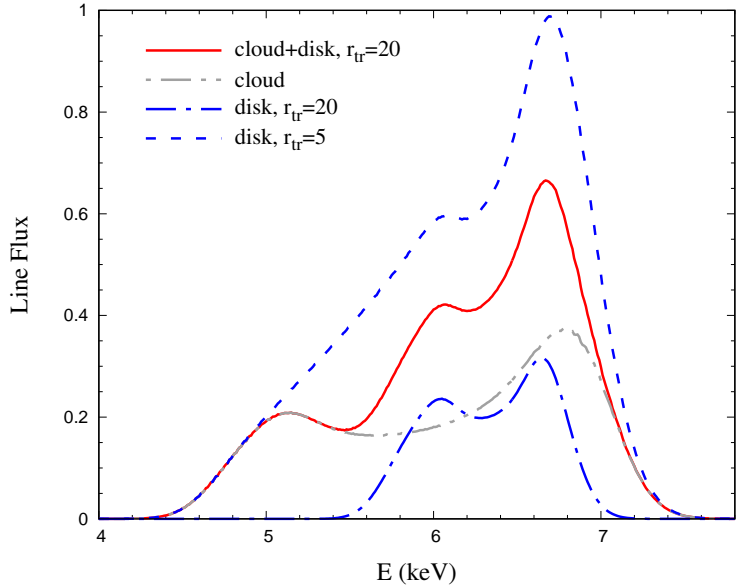}
    \caption{An example of the emission line from a two-phase accretion flow, with the parameters being $a_*=0.9$, $i=30^\circ$, $r_{tr}=20~r_{\rm{ISCO}}$, $r_{cl}=5~r_{\rm{ISCO}}$, and $\Delta r=4~r_{\rm{ISCO}}$. The red solid line shows the total emission line, while the blue dot-dash and the grey dot-dot-dash lines show the components from the cold cloud and the outer cold disk, respectively. For comparison, the emission line of the same width from the continuous disk is also shown by the blue dashed line.}
    \label{fcloud}
\end{center}
\end{figure}

\section{The Model}

\subsection{Geometric structure and justification of the annulus-like clump}

In our phenomenological study, the accretion flow is described with a set of geometric parameters, the variation of which can describe the three scenarios of state transition.
The geometry of the model is shown in Fig.~\ref{fgeom}. Interested readers are referred to \citep{Yu2017,Yu2018} for more details. There are three components in the model. The continuous standard accretion disk lies on the outside of a truncation radius, $r_{\rm tr}$. Inside $r_{\rm tr}$, the disk is replaced by hot gas, either hot accretion flow or corona.
The annulus locates at the mid-plane within the hot gas, representing the cold clumps.
 {Since the inner hot accretion flow is hot and therefore geometrically thick, its X-ray emission can illuminate not only the cold clumps but also the outer cold disk.}

It is necessary to investigate the geometry of the clump. For the problem of state transition, the clump is not far from the central BH, suffering strong tidal force. The tidal force would stretch the clump in radial direction, meanwhile the differential rotation of the accretion flow would stretch the clump into arc form, making the azimuthal size, $\Delta \psi$, relatively large.

For present observations, the integration time of spectral analysis is typically tens to thousands of seconds, in contrast to the orbital timescale of accreting gas of less than a second. In this situation, the clump looks just like annulus to the telescopes except that, the ``observed'' annulus is dimmer, or the iron line flux is lower. The difference of flux directly depends on the intrinsic $\Delta \psi$ for  given radial size and position. Considering the arc form mentioned above, the difference should be within one order of magnitude.
We caution that this result does not hold for AGNs, where the orbital timescale can be minutes to hours, being comparable to the integration time in X-ray spectral analysis.

We additionally note that, there is a degeneracy between the azimuthal size and the radial size. The line strength depends on the area of the cloud. It means that the iron line from a clump of limited azimuthal size is similar to that from a narrower ring at the same radial position. Although the line profiles are a bit different, the difference is too small to be distinguished for present observations. In other words, the radial size and azimuthal size of the clump are degenerate.

Considering the above reasons, it is acceptable to simplify the clump as annulus to study the width-flux relation.
For simplicity, in our calculation we only consider the case with one cloud. As will be shown later, our results are not affected significantly by the number of clumps, because the width and flux of the iron line mostly depend on the radial position of the innermost cloud and the total area of the clumps, respectively.
Therefore the clump can be well described via two parameters, i.e., the radial position, $r_{\rm{cl}}$, and width of the annulus, $\Delta r$. 
 {Different scenarios of the state transition can all be described by the variation of these three parameters.}
For the continuous disk model, the parameters can change in two ways. One is that $r_{\rm{tr}}$ decreases or increases during the state transition, meanwhile $\Delta r=0$ all the time.
The other is that $r_{\rm{cl}}$ decreases or increases during the state transition, meanwhile $r_{\rm{tr}}=\Delta r + r_{\rm{cl}}$ all the time.
For the innermost annulus model, $r_{\rm{cl}}\sim r_{\rm{ISCO}}$ all the time, and $\Delta r$ and $r_{\rm{tr}}$ change with time;
And for the cloud model, the parameters are all free, being limited only by the geometry.

\subsection{Broad skewed iron line profile}

The observed line profiles can be calculated with the ray tracing technique, which has been well developed and widely used \cite[e.g.][]{Cunningham75,RauchBlandford94,Fanton97,Garcia14}. 
We used the method and code developed in \cite{Fanton97}. Briefly, if the centers of the BH and its image are set as the coordinate origins of the disk plane and image plane, respectively, for each grid on the image plane and a given redshift factor, the photon trajectory is determined according to the theory of general relativity \citep{Carter1968}.
It is then possible to trace the trajectory from the image plane backwards to the origin of emission according to the variation of polar angle. If the clumps is assumed to be in the equatorial plane of the BH, the polar angle would change from $\pi/2$ to the inclination angle of eyesight. Considering the axisymmetry of the geometric and the accretion flow, without the need to find the azimuthal position, the radial position where the trajectory intersect with the disk plane can be solved. 

For the obtained radial position, the emissivity and movement can be calculated according to the structure of the accretion flow. Following the usual way of treatment, we assume the emissivity, or the intensity of emission, follows a power-law distribution, i.e., $I\propto r{^{-p}}$ \citep{2012MNRAS.424.1284W,2013MNRAS.430.1694D}, where index $p=3$ is fixed to a typical value. Since the movement of the cloud is still difficult to dynamically constrain, we assume it to be Keplerian just like the outer cold disk. Meanwhile, the solid angle that the image grid subtends to the point of photon emission can be calculated according to the photon geodesics. Subsequently, the observed flux can be obtained with coordinate conversion to the observer. Finally, scan the image plane, and then bin the observed flux by redshift factor, we can obtain the theoretical line profile, as well as the total flux of the iron line.

Here we want to note the assumption we followed, $I\propto r{^{-p}}$. Firstly, the influences of the illuminating X-rays and the reflection rate are not included, in other words, the  illuminating X-rays and reflection rate are normalized in our calculation. Since the geometry parameters of our model are fixed in the soft state, we select the illuminating X-ray flux, the reflection rate, and iron line flux in the soft state as normalization factor, and investigate the relative variation of the iron lines. Secondly, we set the outer boundary of the cold disk at 200~\risco, which does not affect the result obviously.

When the truncation radius is close to the ISCO, the temperature of the cold gas would be as high as about 1~keV, so the broaden of the line on the disk plane due to the thermal Doppler effect should also be taken into account. So we assume the temperature of the cold gas, both the cloud and the outer cold disk, to be that of SAD at the corresponding radius.
The iron line is broadened following Gaussian distribution, $I(\nu)=\frac{1}{\Delta{\nu}\sqrt{\pi}}e^{-\frac{(\nu-\nu_{\rm 0})^{2}}{\Delta{\nu}^2}}$, 
where $\nu_0=6.4~{\rm{keV}}$ is the line frequency in the rest frame, $\Delta\nu=\nu_0\sqrt{\frac{2kT}{m_{\rm{e}}c^2}}$ is the broadening produced by the Doppler effect, $k$ is the Boltzmann constant, $m_{\rm e}c^2$ denotes the rest energy of the electron, and $T$ is the temperature of the cold gas.

\begin{figure}
\includegraphics[angle=-90,width=\columnwidth]{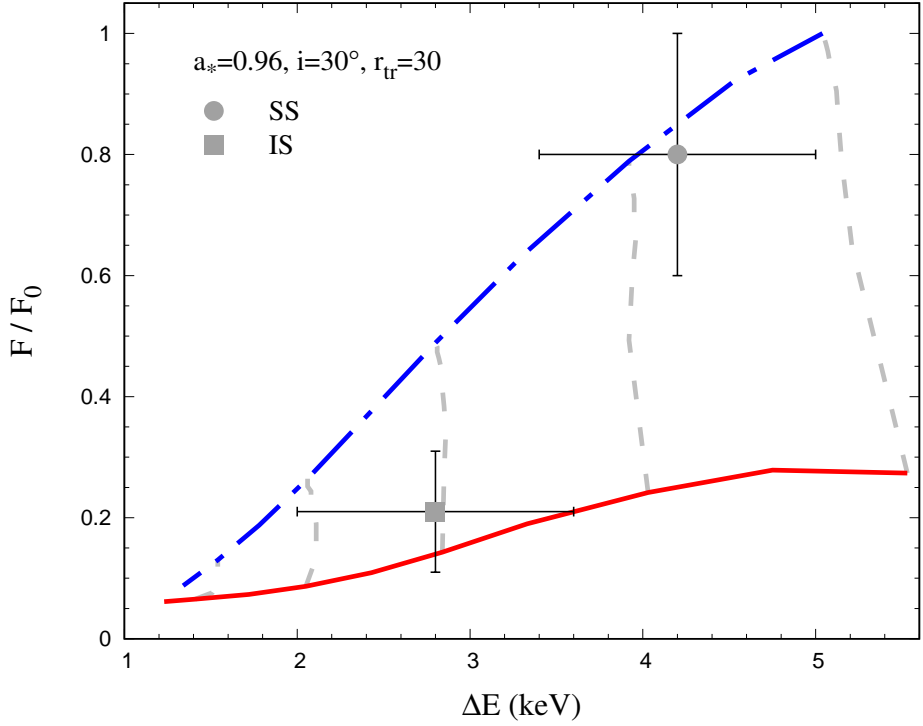}
\caption{The possible width-flux relation during the state transition, where $F_0$ is the line flux from SAD. As example the BH spin and the inclination angle are taken as $a_*=0.96$, $i=30^\circ$. The {\it blue dash-dot} line corresponds to the continuous disk model, or the upper limit of the line flux for given width. The {\it red solid} line corresponds to the relation of the cloud model, where the cloud contribute 10\%  of the total line flux. And the grey dotted lines correspond to the cloud of all possible sizes at given radii, among which the one of highest width corresponds to the innermost annulus model.}
\label{ffw}
\end{figure}

\begin{figure*}
\includegraphics[angle=-90,width=0.32\columnwidth]{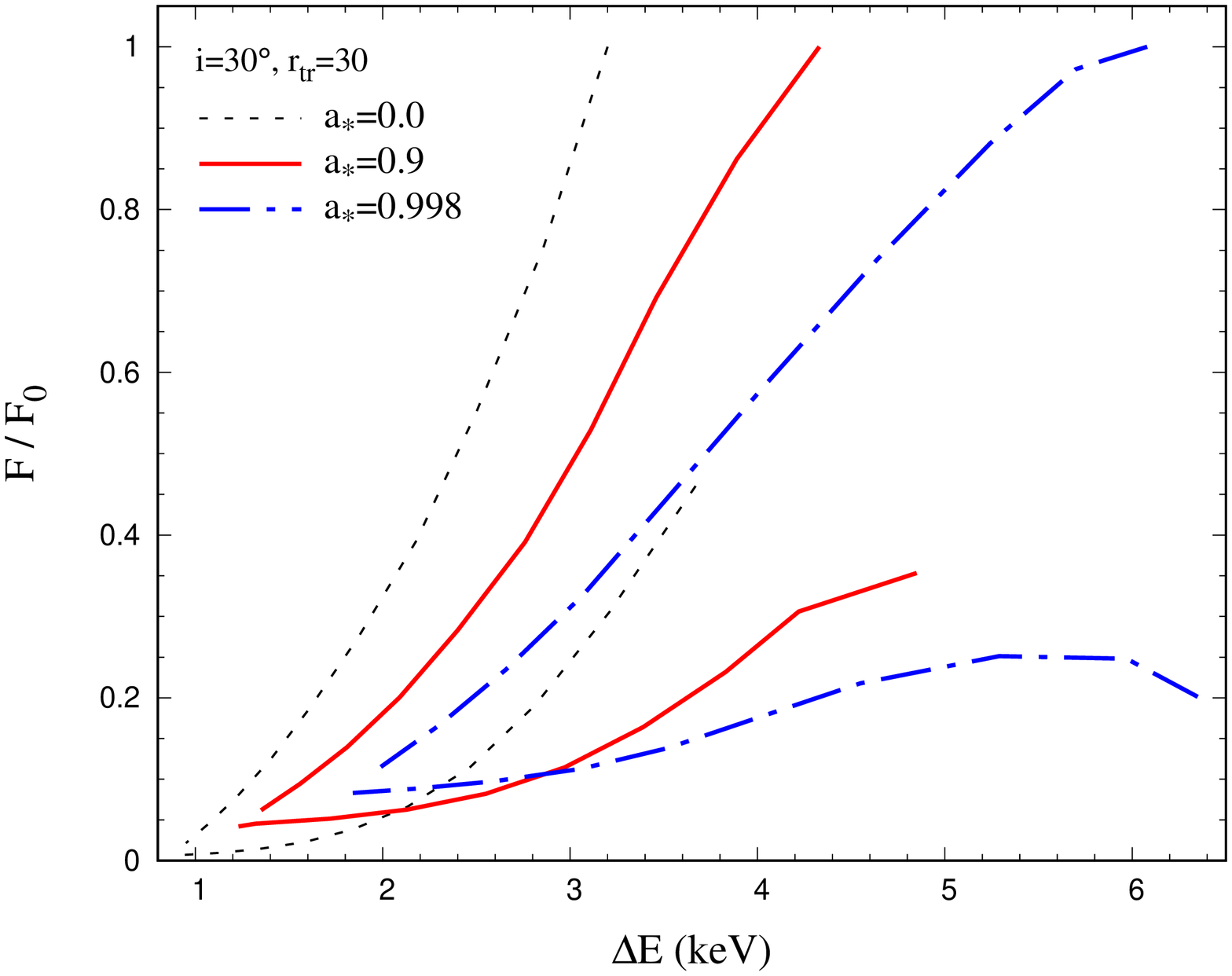}
\includegraphics[angle=-90,width=0.32\columnwidth]{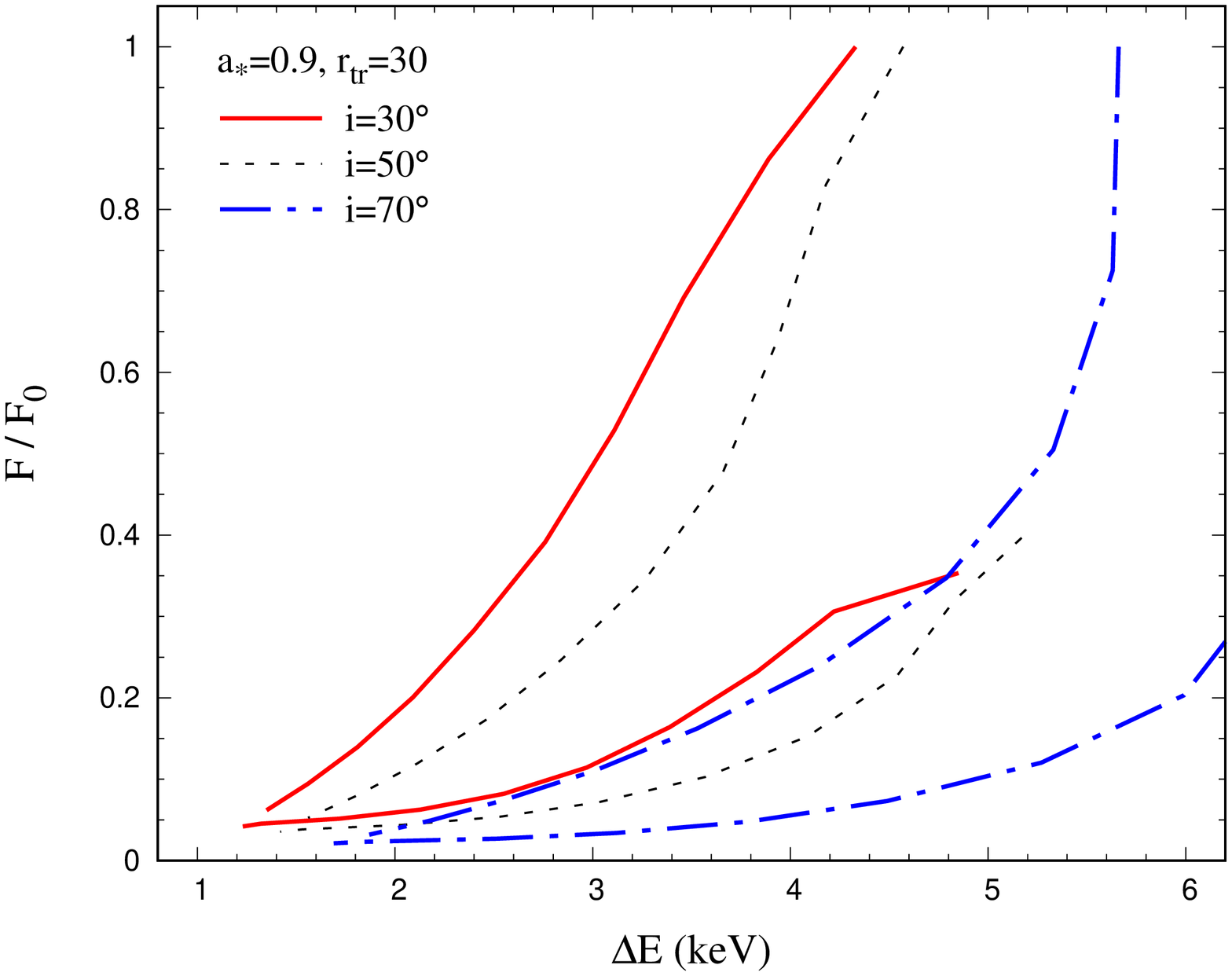}
\includegraphics[angle=-90,width=0.32\columnwidth]{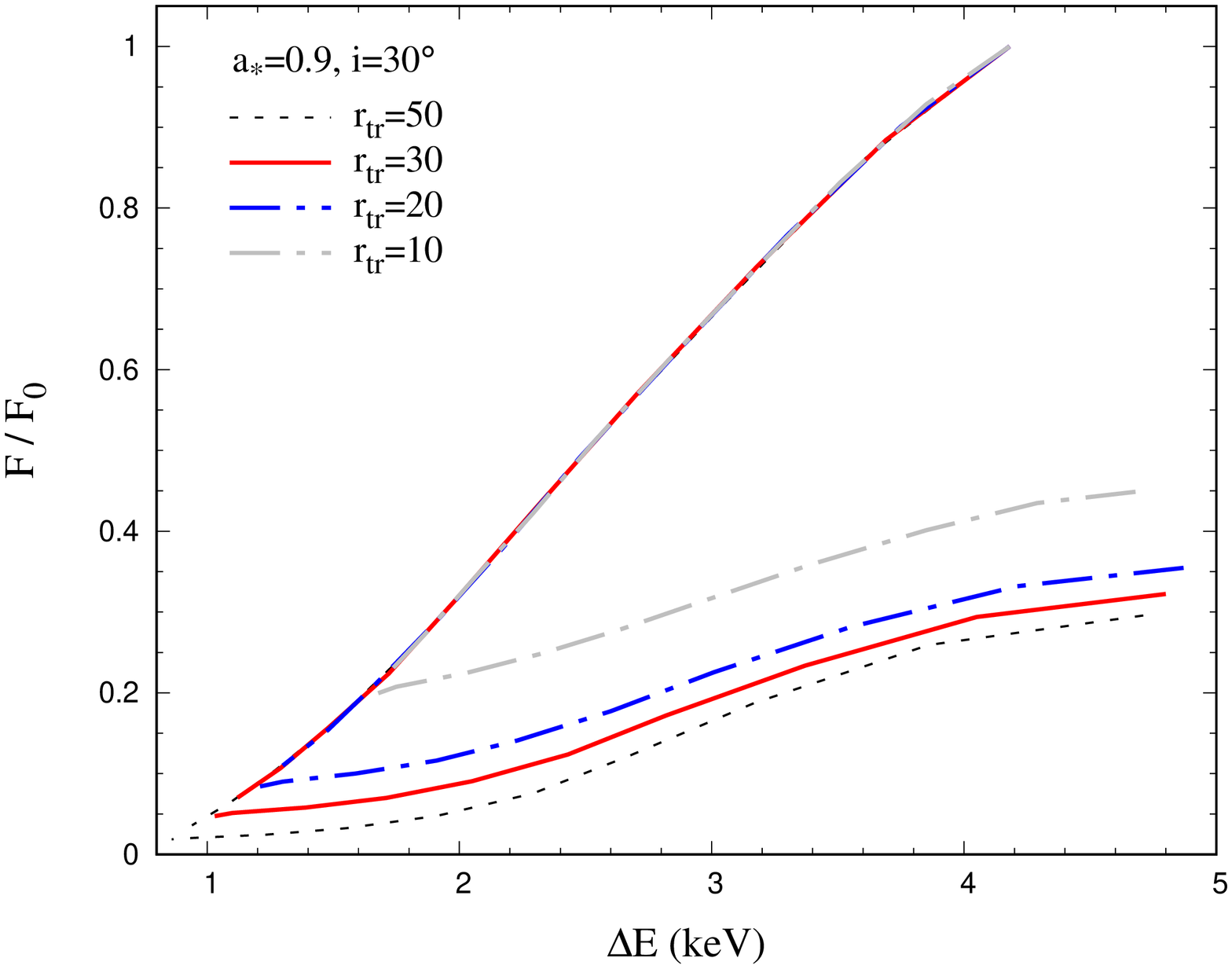}
\caption{The possible width-flux relation for different BH spin ({\it left panel}), inclination ({\it middle panel}), and truncation radius ({\it Right panel}).}
\label{ffr}
\end{figure*}

\section{Results}

In order to show the influence of the clouds on the width and flux of the emission line, Fig.\ref{fcloud} is plotted.
The blue dashed and dash-dot lines show the iron lines emit from the outer cold disks truncate at $5~r_{\rm ISCO}$ and $20~ r_{\rm ISCO}$, respectively.
The grey dot-dash-dot line is the contribution of a cloud at $5~r_{\rm ISCO}$.
The iron line from the two-phase accretion flow, in which $r_{\rm tr}=20~r_{\rm ISCO}$ and $r_{\rm cl}=5~r_{\rm ISCO}$,
is the sum of the dash-dot and dot-dash-dot lines.
As can be seen that the width of the line depends most on the inner boundary of the cold gas.
Therefore, it is a natural result that for the given line width, the flux of the continuous disk is the highest, since the area of reflection is largest.

For a given BH spin and inclination angle, as the geometry of the cold gas evolves, the flux and width of the emission line vary correspondingly.  Because the geometric parameters change in different ways, the relation between the flux and width differs for different scenarios. The range of possible relations are shown in Fig.\ref{ffw}.
It should be noted that the width we used is defined by 10\% of the peak flux instead of half maximum, otherwise a large fraction of the red wing would be ignored.
Moreover, the line flux is normalized by the iron line in SS that is contributed by the complete area of SAD, as mentioned before. Compare to the absolute flux, the normalized flux shows the relative variation.

Considering an annulus at $r_{cl}$, when $\Delta r$ increases, the total line flux increases as the area of reflection becomes larger, meanwhile the line width remain almost the same since the red wing of the line profile mostly depend on the inner edge of the ring, where the gravitational redshift is maximal. 
The influence of $\Delta r$ is shown by the grey dashed lines. The first one from right represents the width-flux relation for the innermost annulus scenario.
The blue dash-dot line indicates the highest flux, which correspond to the largest cloud that fill up all the area between $r_{\rm cl}$ and $r_{\rm tr}$. In other words, it represent the relation for the continuous disk scenario.
The red solid line shows the lowest line flux that corresponds to the most narrow ring we set.
Theoretically, the line flux from the annulus could be as small as zero.
But from the view point of observation, the flux from the cloud cannot be too small, otherwise the contribution by the cloud would be undetectable. 
Here we set the most narrow ring be about 10\% of the gap region between $r_{cl}$ and $r_{tr}$. 
Generally speaking, the width-flux relation should be within the region between the blue dash-dot line and the red solid line.

The influence of the BH spin on the range of the line flux is shown in the left panel of Fig.\ref{ffr}, in which the upper and lower limits are plotted with the same type of line as shown in the caption.
When the spin increases, the maximal width becomes larger as the ISCO is closer to the BH and the gravitational effects and the Doppler effect are more significant.
It can be found that for the cases with broader lines, the range of the flux becomes larger with increasing $a_*$.
This is because the broader lines are produced by cold gas closer to the BH, which suffers more gravitational redshift.

Similarly, the influences of the inclination angle and the truncation radius are shown in
the middle and right panels of Fig.\ref{ffr}, respectively.
When the inclination angle becomes larger, or the line of sight is closer to the equatorial plane, the width of the line increase significantly due to the Doppler effect.
Therefore the width-flux relation is very sensitive to the inclination.
For larger truncation radius the contribution of the outer cold disk is smaller, as a consequence the lower limit, i.e. 10\% of the peak flux becomes smaller, which means the cloud could be smaller. And so it is easy to understand why the lower limit decreases with increasing truncation radius.


\section{Application to \target}

Although our model is very simple, it provides us a quantitative and case independent results. Based on the spectral fitting of the observational data, the spectra and intensity of the illuminating X-rays, as well as the degree of ionization, can be obtained. It is then possible to estimate the normalized iron line flux by use of the relation between reflection and ionization level \citep[e.g.][]{Ross1999}.

\subsection{Previous observational results of \target}
The black hole X-ray binary \target was discovered in 2018 by the All Sky X-ray Image Monitor \citep[MAXI,][]{2018ATel12320....1K}. And after that many interesting works have been done on this source.
\cite{Xu_2020} carried out a spectroscopic analysis based on the observation of the Nuclear Spectroscopic Telescope Array (\NuSTAR) during its 2019 outburst.
They captured two epochs during the decay of the 2018-2019 outburst and clearly detected the strong relativistic disk reflection features in both SS and IS.
\cite{2021MNRAS.505.1213R} performed a combined spectral and timing analysis by use of the observational data by the Neutron Star Interior Composition Explorer (NICER),
and \cite{2021ApJ...919...92B} presented the broadband timing properties from the Insight-Hard X-ray Modulation Telescope (Insight-HXMT) observations.

\begin{figure}
 	\includegraphics[angle=-90,width=\columnwidth]{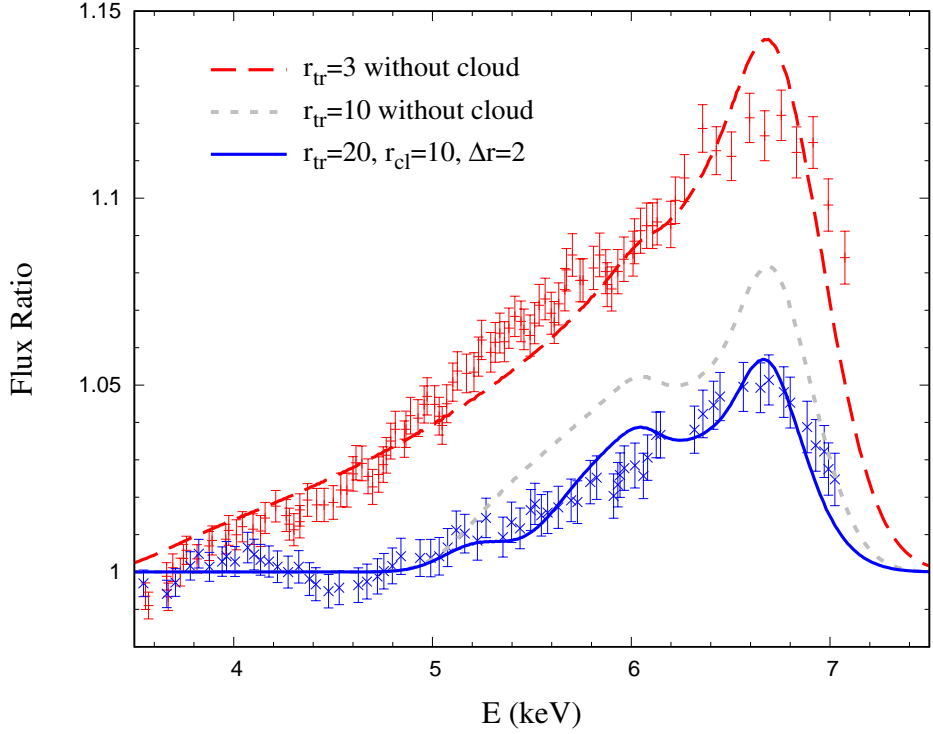}
    \caption{The modeling of the broad emission line from  \target observed by \NuSTAR. The observational data are taken from \citet{Xu_2020}, with the red and blue dots correspond to Observation IDs 90501301001 (Obs1 part I) and 80401316004 (Obs3), respectively. The data above 7 keV of all observations are ignored. The blue solid line is the modeling with the cloud model. The long dashed line are the modelling to the line in SS with the continuous disk model. The short dashed line is the modeling to the line in IS with continuous disk model, which failed to explain either the width or the strength without additional variation of observational parameters.}
    \label{spob}
\end{figure}

The observations of \NuSTAR boast of the broad passband (3-79 keV) and the high spectral resolution.
Being free from the pile-up effect, they are well suited to study relativistic reflection and Fe K$\alpha$ emission line.
So in this work we concentrate on the result by \cite{Xu_2020}.
The authors separated the two epochs of observation into four stages (Obs 1 part I and II, Obs 2, and Obs 3).
The first two (obs1 part I and part II) are obtained from the observation of ID 90501301001, and the spectra are dominated by thermal component, corresponding to the SS.
The rest two stages (Obs 2 and 3) are from the observations of ID 80401316002 and 80401316004, and the spectra are dominated by the power-law component, corresponding to the IS.
After subtracting the continuum, i.e., thermal and power-law components,  they found excess in the 5-7~keV, or the iron line profile as shown in Figure 3 of their paper. 

Considering the iron line profile is relatively constant over shorter intervals, we select the Obs1 part I and Obs3 to represent the disk dominant and power-law dominate states. 
 {Here we summarize the related results from \citet{Xu_2020}. The spin of \target to be $a = 0.96$ and the inclination angle to be $i = 30^\circ$. When the source evolves from part I of Obs1 to Obs3, the line width varies from about 3 keV to 2 keV, the line flux decreases from about $8\times 10^{-10}~$\ergscm to $2\times 10^{-10}~$\ergscm, the illuminating X-ray flux (10-79~keV) increases from about $10^{-9}~$\ergscm to $10^{-8}$\ergscm, the illuminating X-ray flux (3-10~keV) reduces from about $2.5\times10^{-8}~$\ergscm to $2.3\times10^{-8}$\ergscm, the photon index hardens from about 2.5 to 2.4, the ionization parameter, $\xi$, decreases from about 60000~\ergcms to 3000~\ergcms, and the thermal and power-law components of the continuum are all fitted.}

\subsection{Width-flux relationship in \target}

For the given spin and inclination angle, the theoretical line width could be as high as about 5~keV for SAD when $a = 0.96$ and $i = 30^\circ$, while as the fitting results of the observation is about 3.4~keV. The difference of 1.6~keV is taken to be the  error range of the line width for consistence. For obs3, the theoretical line width can not be determined, as the geometry of the accretion flow is unknown. So we also assume it to be 1.6~keV. 

As the line flux is estimated with a Gaussian line in \cite{Xu_2020}, we assume an error of 0.2 dex, which means the possible range is about 40\%.
So the possible region of obs1 in the width-flux plane can be limited, the center of which is marked with solid circle point.

The relative line flux in IS is estimated as folows. 
According to the fitted continuum, the total illuminating X-rays in the range of 7-20~keV, which is most important in ionizing electrons in the innermost shell of iron, increases by about 3 times in IS.  
To estimate the variation of reflection rate for different ionization level, we used the ``relxill" model in XSPEC to obtain the reflection spectra and then compare the fluxes in 6-7~keV for the illuminating X-ray spectra ($\Gamma=2.4$) and ionization degree ($\xi=10000~and~3000$~\ergcms) corresponding to SS and IS, respectively. We found that for the same intensity of illuminating X-rays, the iron line flux decreased by about 2 times in IS. To compare with our theoretical relation, the line flux should be normalized to the same intensity of illuminating X-rays and the same reflection rate. Therefore the ratio of the line flux of IS to that of SS should be further divided by 3 and multiplied by 2, which is finally about 0.17. Considering the 40\% uncertainty on the line flux, the possible range of normalized flux is 0.10$\sim$0.22, as shown by the solid square point.

Then we can compare the observation results with the our theoretical results. 
As shown in Fig.~\ref{ffw}, the normalized line flux for the soft state roughly agrees with the theoretical result, or the continuous disk model. While the data point for the intermediate state located below the continuous disk model, but can still be well explained with the cloud model.

\subsection{Modelling of the iron line profiles}

To further check the result, instead of just considering the width and strength, we modeled the variation of the line profile by changing the parameters of the cold gas reasonably, as shown by the curves in Fig.\ref{spob}.
The line profile in the soft state can be modeled with a cold disk that extends to the boundary of 3 $r_{\rm ISCO}$.
For the intermediate state, the truncation of the cold disk occurs at about 20 $r_{\rm ISCO}$, meanwhile the cold gas inside the truncation radius is in the range of 10 - 12 $r_{\rm ISCO}$. Our result agrees with that of \citet{Xu_2020}, as $r_{\rm ISCO}$ is close to gravitational radius for fast spinning BHs. 

We also tried to model the variation of the line profiles with the continuous disk model by just changing the truncation radius, we find it is difficult to model the profile when take into account the variation of the illuminating hard X-rays and ionization degree. As show by the short dashed line in Fig.~\ref{spob} for example, this model failed to explain the line strength and width at the same time.

 {Moreover, it is interesting to explore the impact of the azimuthal size of the cloud. If the azimuthal size at a given radius for each cloud is less than $30^\circ$, which is about one-tenth of $2\pi$, the radial range would be at least ten times larger, considering the degeneracy of the azimuthal and radial size. If there is just one cloud, it should extend from $10~r_{ISCO}$ to $30~r_{ISCO}$, which is obviously unreasonable because the outer boundary of the cloud is larger than the truncation radius. So we expect that there should not be just one cloud.  As the emissivity of the iron line is decreasing with radius, we expect that the number of the cloud should be at least 3.
However, according to the evaporation model of the corona, where the evaporation rate depends on the radius, it is still possible for a annulus to form during the state transition.}

\section{Summary}

In this work, we investigated the variation of the width and strength of the broad iron lines during the state transition, and explored the possible process of state transition. 
When compare with the combined observational data of \target, the cloud model is favored according to the width-flux relation we obtained.
Moreover, the cloud model can explain not only the spectra but also the relatively high reflection rate in SS.
If the azimuthal size of the cloud is small, our modeling parameters indicate that the number of cloud should be more than 2.

In this work, the line flux and width in SS is selected as the normalization. It is necessary to further check the variation of the iron during SS. Moreover, the ionization level is assumed to be the same for the clouds and outer cold disk, we would further improve the calculation in future works.

\section*{Acknowledgements}
This work was supported in part by the National Natural Science Foundation of
China under grant number U2038108, 12192220, 12192223 and 12133008, the National SKA Program of China (No.2020SKA0110102). F.G.X. is additionally supported in part by the Youth Innovation Promotion Association of CAS (Y202064).








\label{lastpage}
\end{document}